\documentclass[12pt]{article}

\usepackage{amsmath,amsfonts,amssymb}
\usepackage{graphicx}
\usepackage{supertabular}
\usepackage{fancyhdr}
\usepackage{placeins}
\begin{document}

\title{An Indel-Resistant Error-Correcting Code for DNA-Based Information Storage}

\author{W.H.~Press and J.A.~Hawkins\\ Institute for Computational
  Engineering and Sciences\\
  University of Texas at Austin}
\maketitle

\newcommand{\onebf}{\boldsymbol{1}}
\newcommand{\Deltabf}{\boldsymbol{\Delta}}
\newcommand{\prob}{\text{prob}}
\newcommand{\mystrut}{\vphantom{\Bigg)}}
\newcommand{\argmax}{\operatornamewithlimits{argmax}}
\newcommand{\z}{\mathbb{Z}}
\newcommand{\HHH}{H\mathrel{\mathop:}=}
\newcommand{\HH}{H\text{:=}\,}
\newcommand{\Pdel}{P_{\text{del}}}
\newcommand{\Pins}{P_{\text{ins}}}
\newcommand{\Psub}{P_{\text{sub}}}
\newcommand{\Pok}{P_{\text{ok}}}
\newcommand{\Peq}{P_{\text{equiv}}}
\newcommand{\peq}{p_{\text{equiv}}}
\thispagestyle{fancy}

\section{Introduction}
Engineered DNA is an information channel.  One can convert an
arbitrary message into a string of DNA characters, or bases,
$\{A,C,G,T\}$, synthesize the string into a physical DNA sample; store
or transport the sample through space and time; sequence it back to a
string of characters; and then hope to recover exactly the original
message.  Because errors are introduced during all the stages of
synthesis, storage, and sequencing, it is necessary to utilize an
error-correcting code (ECC) at the stage of converting message bits to
DNA characters (encoding), and then later, when DNA characters are
converted back to message bits (decoding).  The ECC needs to correct
three kinds of errors: substitutions of one base by another,
spurious insertions of bases, and deletions of bases from the message.
Insertions and deletions are commonly termed ``indels''.

The correction of substitutions is a standard problem in coding
theory, where substitutions are termed ``errors''.  The overarching
theoretical framework for coding theory starts with Shannon
\cite{sh}, and there exist hundreds, if not thousands, of well studied
error-correcting codes (ECCs) \cite{macwilliams,roth,moon,lin}.
However, established methods for error correction in the case of
silent deletions---termed deletion channels---are few; and there are
virtually no established methods for channels with all three of deletions,
insertions, and substitutions.  (See \cite{mitz} and \cite{rayli} for
reviews and references.)  Indeed, no approaches suggested in the
literature are well suited to DNA applications \cite{barcodes}. For
example, almost all attention has been on binary channels, while the DNA
channel is quaternary.

As the main contribution of this paper,
we describe in Section \ref{theCode} a method for encoding a stream of
arbitrary message bits onto a stream of DNA characters with an ECC
that simultaneously corrects all three kinds of errors.  Our ECC,
which we call HEDGES (for ``Hash Encoded, Decoded by Greedy Exhaustive
Search''), is tuned to recover character-by-character message
synchronization, even at the cost of leaving a small number of
uncorrected substitution errors .  This tuning makes HEDGES useful as
the ``inner'' code (closest to the channel and applied last in
encoding) in a concatenated code design, leaving it to a conventional
``outer'' code (applied first in encoding, last in decoding) to
correct any remaining substitution errors.  Below, in combination with
HEDGES, we will use the standard Reed-Solomon (R-S) code \cite{wicker}
denoted RS(255,223).  R-S codes are completely intolerant of indels,
i.e., they require perfectly synchronized input.

Because of our tuning of HEDGES for use in a concatenated design, it
will be useful to first describe a possible full system design
(Section \ref{systemdesign}), and only then describe HEDGES in detail
(Section \ref{theCode}).  The system design in Section
\ref{systemdesign} is illustrative but not unique.  HEDGES itself, as
a quaternary-alphabet indel-resistant ECC, is general and can readily
be utilized in other overall designs.

\section{Related Work}\label{relatedwork}

There is a growing body of experimental work on DNA information
storage, employing various strategies for dealing with errors.  We
summarize chronologically some of the previous work as it relates to
this paper.

Church et.~al \cite{churchgao} synthesized oligomers of length 159,
each of which contained both address information (ordering of
oligomers in the message) and payload.  There was no explicit ECC.
The pool of oligomers was sequenced to a depth 3000x, allowing recovery
of a consensus sequence with high probability.  High-depth coverage
can correct sequencing errors, but not synthesis errors.  Indeed, the
final results contained 10-bit errors.

Goldman et.~al \cite{goldman} synthesized oligos with 3/4 of each strand
overlapping the previous strand, in effect a 4x repetition code.  Each
strand had parity check bits for error detection (but not correction).  A
ternary code (ternary message alphabet to quaternary DNA alphabet)
was utilized to avoid homopolymers.  Error correction was done by
sequencing to high depth and filtering for, and aligning, perfectly
sequenced fragments.  The final results contained two gaps where none
of the four overlapping sequences were recovered.

Grass et.~al \cite{grass} implemented interleaved Reed-Solomon codes for DNA
storage.  Message bits were converted to characters in the
47-character alphabet GF(47), with interleaved correction in blocks of
$(713\times 39)$ characters. An inner code mapped GF(47) characters to
47 DNA trimers chosen to guarantee no homopolymers of length $>3$ in
the final DNA message, but with no other redundancy.  Indels were
corrected by sequencing at sufficient depth to reject faulty strands.

Bornholt et.~al \cite{bornholt} introduced strand-to-strand redundancy
by creating strand $C = A \oplus B$ (or other redundant combinations),
where $\oplus$ denotes exclusive-or, and utilizing majority dedoding.
Parity check bits allowed the filtering out of faulty strands.  There
was no explicit correction of indels.

Erlich and Zielinski \cite{erlich} utilized quite a different overall
architecture, based on fountain codes. Fountain codes send linear
combinations of portions of a message in ``droplets'', such that one
can recover the original message by the solution of linear equations.
The included redundancy allows the loss of some droplets.  R-S coding
was used within each oligomer droplet for error detection but not
correction.  Faulty droplets, including any with indels, were
rejected.

Yazdi et.~al \cite{yazdi} leveraged the multiple sequence alignment
capabilities of several sophisticated packages in conjunction with a
custom homopolymer check code to correct the large error
rates---especially homopolymer errors---associated with MinION
nanopore sequencing.  Sequencing depths were in the range of several
hundred.

In the largest-scale experiment to date, Organick et.~al
\cite{organick} encoded and recovered, error-free, more than 200 MB of
data.  R-S coding was used across strands, with no explicit error
correction within a strand.  Substitutions and indels within a strand
were corrected by multiple alignment and consensus calling. Coverage
was 5x for high-quality Illumina sequencing, rising to 36x to 80x
required for Nanopore technology.

To summarize, while previous work has adopted increasingly
sophisticated system designs, there has been little progress in the
fundamental problem of correcting indels within a single strands.  The
use of sequencing to large depth, followed by multiple alignment, is in
effect a use of the oldest, simplest, and arguably least efficient
ECC, namely a simple repetition code, sending the same message
multiple times and taking the consensus.  This manifests itself as
sequencing stored DNA to high coverage, finding sets of reads which
appear to derive from the same intended sequence, and either merging
the reads into a consensus sequence and/or filtering out any reads
which fail some quality check.  Though this is a central part of
previously implemented DNA error-correction schemes, it also tends to
be left unaccounted for in claims for code efficiency.
The central result in this paper is a technique for correcting
substitutions and indels in a single strand, i.e., when sequenced to
no more than depth one.

\section{Example System Design}\label{systemdesign}

\begin{figure}[t]
\centering
\includegraphics[width=6.25in]{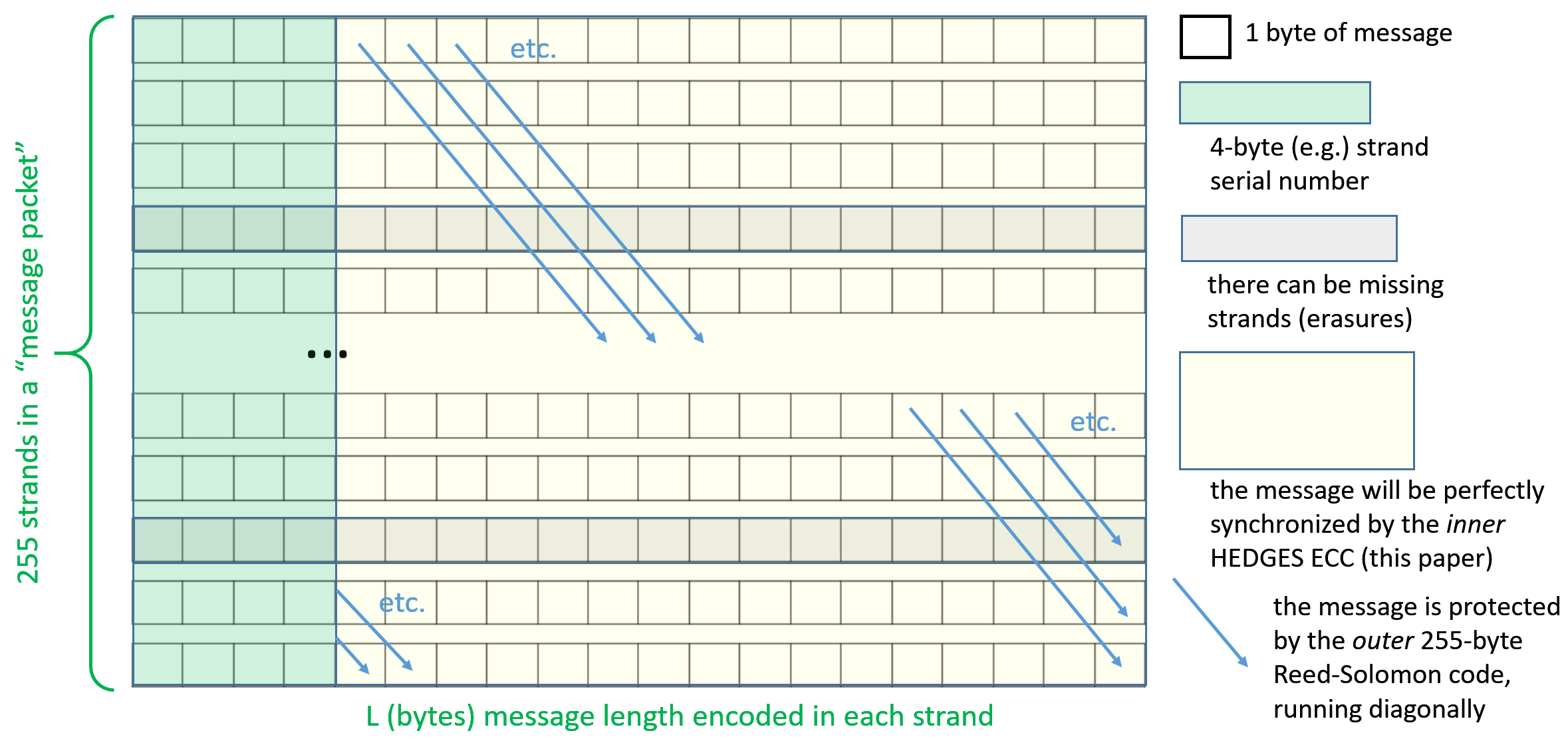}
\caption{The overall system design is a concatenated code with
  interleaving across 255 strands of DNA (horizontal lines in the
  figure).  Each strand is protected by HEDGES, this paper's
  indel-correcting code.  HEDGES restores synchronization, so that the
  packet can then be protected against residual errors and missing
  strands by a Reed-Solomon code (applied diagonally, as shown, for
  reasons described in the text).}
\label{fig1}
\end{figure}

For cost and efficiency, both DNA synthesis and DNA sequencing employ
massive parallelism.  That is, many short sequences, each of length
hundreds to thousands of bases, are written (synthesized) or read
(sequenced) simultaneously.  While the length of a single synthesis or
read will increase as technology improves, it is unlikely that the great
advantage of parallelism will ever be superceded.  This being the
case, the basic units of our system design are individual strands of
length $10^2$--$10^4$.

To connect with our use of RS(255,223), we define a ``DNA packet'' as
an ordered set of 255 DNA strands.  When any one strand in the set is
decoded with HEDGES, it produces a message fragment of length $L$
bytes (say), now having high probability of being perfectly
synchronized.  Each 255 correctly ordered message fragments form a
``message packet'', as illustrated in Figure \ref{fig1}.  There can
be any number of message packets in a total communication.

The Reed-Solomon code is applied across the strands (interleaved).
This enables it to protect against missing strands---``erasures'' to
coding theorists---as well as correcting any residual substitution
errors that were not corrected by HEDGES.  Different from previous
investigations, we apply the R-S code diagonally across the strands
(see Figure).  This increases the resistance to any failure of
synthesis or sequencing to produce full-length strands.  It also
ameliorates the effect of the observed tendency for error rates
to be higher at the ends of strands.

It is an important point that the Reed-Solomon code can only be
applied after the strands in a packet are identified as being from one
particular packet (out of an assumed pool of many packets, perhaps
millions) and are correctly ordered.  This implies that a packet's
identification number and a strand's serial number within the packet
(both shown as shaded green in the Figure) cannot themselves be R-S
protected.  We will instead protect them by a different technique
(``salt protection'') that is described in Section \ref{saltsec}.
Salt protection has the effect of turning uncorrectable errors in the
identification/serial bytes into erasures in the message bytes---which
are correctable by R-S.

Summarizing, here are the main points that affect the design of HEDGES
as an inner code: (1) We don't need to decode strands of arbitrary
length, but only of some known uncorrupted length $L$.  (2) Recovering
synchronization has the highest priority.  (3) Known erasures are less
harmful than unknown substitutions, because R-S can correct twice as
many erasures as substitution errors.  (4) Burst errors within a
single byte are less harmful than distributed bit errors, because R-S
corrects a byte at a time.  (5) Within the R-S code's capacity for
byte errors and erasures, residual errors will be fully corrected by
the outer code, yielding an error-free message.

\section{HEDGES, an Indel-Correcting Code}\label{theCode}

\subsection{Overall Strategy}

Given a message stream of bits
\begin{equation}
  b_i,\quad i=0,1,2,\ldots,M,\quad b_i \in \{0,1\}
\label{bitstream}\end{equation}
(``the message'' or ``bits''), we want to emit a stream of DNA characters
\begin{equation}
C_i,\quad i=0,1,2,\ldots,N,\quad C_i \in \{A,C,G,T\} \equiv \{0,1,2,3\}
\notag\end{equation}
(``the codestream'' or ``characters'').  We first describe the case of
a half-rate code, where we emit exactly one $C_i$ (2 bits of output)
for each $b_i$ (1 bit of input).  In section \ref{othercoderates} we
generalize to codes at other rates $r$ (message bits per codestream
bit), $0 < r < 1$, so that the streams $b_i$ and $C_i$ are not then in
lockstep, and $M \ne N$. One should think of $N$ as being on the order
of $10^2$ to $10^4$, the maximum length of a single DNA strand that
can be cheaply synthesized today or in the foreseeable future.

We want to be able to decode without residual errors a received
codestream $C^\prime$ that differs from $C$ by substitutions (errors),
insertions, and deletions (collectively ``indels'').
Indels are silent: their positions in the codestream $C^\prime$ are
not known to the receiver.

The basic plan is a variant of a centuries-old cryptographic
technique, ``text auto-key encoding'' \cite{vi}.  We generate a
keystream of characters $K_i \in \{0,1,2,3\}$, where each $K_i$
depends pseudorandomly (but deterministically by a hash function) on
some number of previous message bits $b_j$ (with $j<i$), and also
directly on the bit position index $i$.  (We can initialize the
previous bits by defining $b_j \equiv 0$ when $j<0$.)  We then emit a
codestream character $C_i = K_i + b_i$, the addition performed modulo
4.  In the terminology of modern code theory, this scheme would be
called a type of ``tree code'' or, more specifically, an ``infinite
constraint-length convolutional code''.

The redundancy necessary for error correction comes from the fact that
$b_i$ takes on only two values, while $K_i$ and $C_i$ can have four
values.  This generates (only) one bit of redundancy per character,
i.e., can be acausally valid by chance half the time.  However, the
dependence of $K_i$ on many previous message bits ties any given
message bit to many future bits of redundancy.  Similarly, the
dependence of $K_i$ on $i$ ties every bit to its position index, so
that (as we will see) insertions can be identified and removed, and
deleted values can be restored.

What is not obvious is that a codestream thus generated can actually
be practically decoded, especially in the presence of errors and
indels at significant rates.  We will show by numerical simulation
that it can be, remarkably easily, essentially by guessing successive
message bits and scoring against the likelihood of the codestream
under the guessed hypothesis.  Wrong guesses will be rejected by
implying exponentially small downstream likelihoods.  In coding
theory, this general technique is known as ``sequential decoding''.

\subsection{Encoding Algorithm}\label{encoding}

Elaborating slightly on the above description, let $S_i$ denote an
arbitrary $s$-bit value (``salt'') that can depend on $i$ but is known
to both sender and receiver,
\begin{equation}
S_i = \text{known} \in \z_2^{\otimes s} 
\notag\end{equation}
Denote the low-order $q$ bits of
the bit position index $i$ by
\begin{equation}
  I_i \equiv i \;(\text{mod } 2^q)
\notag\end{equation}
Let $B_i$ denote the $r$ previous concatenated bits
\begin{equation}
B_i \equiv [b_{i-r}b_{i-r+1}\cdots b_{i-1}] \in \z_2^{\otimes r} 
\notag\end{equation}

\begin{figure}[t]
\centering
\includegraphics[width=6in]{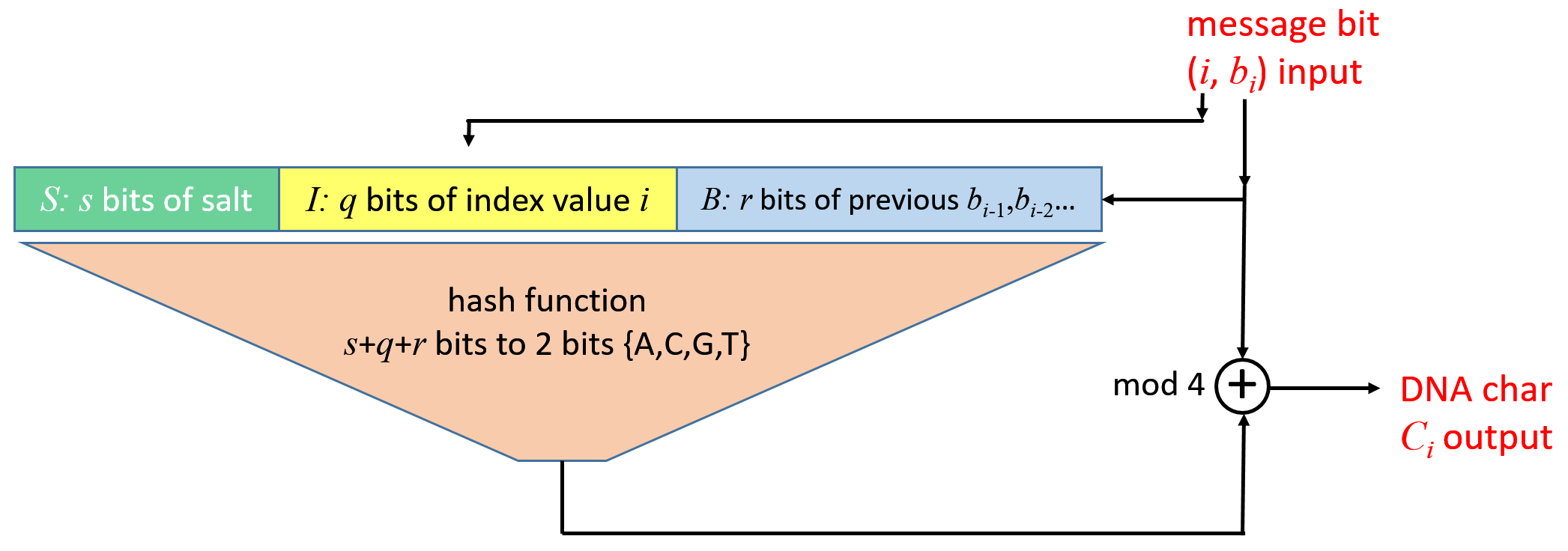}
\caption{The basic HEDGES encoding algorithm is a variant of plaintext
  auto-key, but with redundancy introduced because (in the case
  of a half-rate code, e.g.) one bit of input
  produces two bits of output.}
\label{fig2}
\end{figure}

\noindent Finally, let $F(S,I,B)$ be a deterministic hash function from $r+q+s$
bits to 2 bits
\begin{equation}
F(S,I,B):\quad \z_2^{\otimes (r+q+s)} \rightarrow \z_4
\notag\end{equation}
Then the formula for encoding is
\begin{equation}
C_i = K_i + b_i = F(S_i,I_i,B_i) + b_i \quad(\text{mod } 4)
\label{encodeeq}
\end{equation}
Figure \ref{fig2} shows the algorithm graphically.

Typical values that we use are $r=8$, $q=10$, $s=46$, so that $r+q+s =
64$ bits, a convenient value for input to the hash. For the hash
function we use the low order 2 bits from the {\em Numerical
  Recipes} \cite{nr} function {\tt Ranhash.int64()}, because it is
very fast and will occur in the inner loop of the decode algorithm.

\subsection{Decoding Algorithm}\label{decoding}

For simplicity, assume that error rates are ``small'', so that
``most'' DNA bases are received as they were intended.  (We will see
in Section \ref{validation} that DNA character error rates up to
$\sim5$\%--$10$\% are tolerable.)  Suppose we have correctly decoded
and synchronized the message through bit $b_{i-1}$ and now want to
know bit $b_i$.  Guessing the two possibilities, $\{0,1\}$, we use
equation \eqref{encodeeq} to predict two possibilities for the
character $C_i$.  In the absence of an error, only one of these is
guaranteed to agree with the observed character $C_i^\prime$.  We
assign to a guess that generates disagreement with $C_i^\prime$ a
penalty score equal (conceptually) to the negative log probability of
observing a substitution error.  In other words, a wrong guess might
actually be right, but only if a substitution has occurred.  If
neither guess produces the correct $C_i$, then both are assigned the
substitution penalty.

We have not yet accounted for the possibility of insertions and
deletions, however.  In fact, there are more than the above two possible
guesses.  We must guess not just $b_i \in \{0,1\}$, but also a
``skew'' $\Delta \in \{\ldots,-1,0,1,\ldots\}$ that tells us whether
in comparing $C$ to $C^\prime$
we should skip characters ($\Delta > 0$) because of insertions, or
posit missing characters ($\Delta < 0$) because of deletions
(in which case there is no comparison to be done).  As a
practical simplification we consider only $\Delta \in \{-1,0,1\}$. (We
comment on this simplification in Section \ref{tradeoffs}.)  Then there are six guesses for
$(b_i,\Delta) \in \{0,1\}\otimes \{-1,0,1\}$. Each can be scored by an
appropriate log probability penalty for any implied substitution,
insertion, or deletion.

Log probability penalties accumulate additively along any chain of
guesses.  In the causal case of a chain of all-correct guesses, we
accumulate penalties only in the (relatively rare) case of actual
errors.  However, because of the way that the key $K_i$ (equation
\eqref{encodeeq}) is constructed, single wrong guess for either $b_i$,
$i$, or $\Delta$ throws us into the acausal case where 3/4 of
subsequent comparisons of computed $C$ (at some bit position index $i$) to
observed $C^\prime$ (at some index $k$) will not agree---thus
penalties will accumulate rapidly.  The decoding problem, conceptually
a maximum likelihood search, thus reduces to a shortest-path search in
a tree with branching factor 6, but with the saving grace that the
correct path will be much shorter than any deviation from it.

\begin{figure}[thb]
\centering
\includegraphics[width=4in]{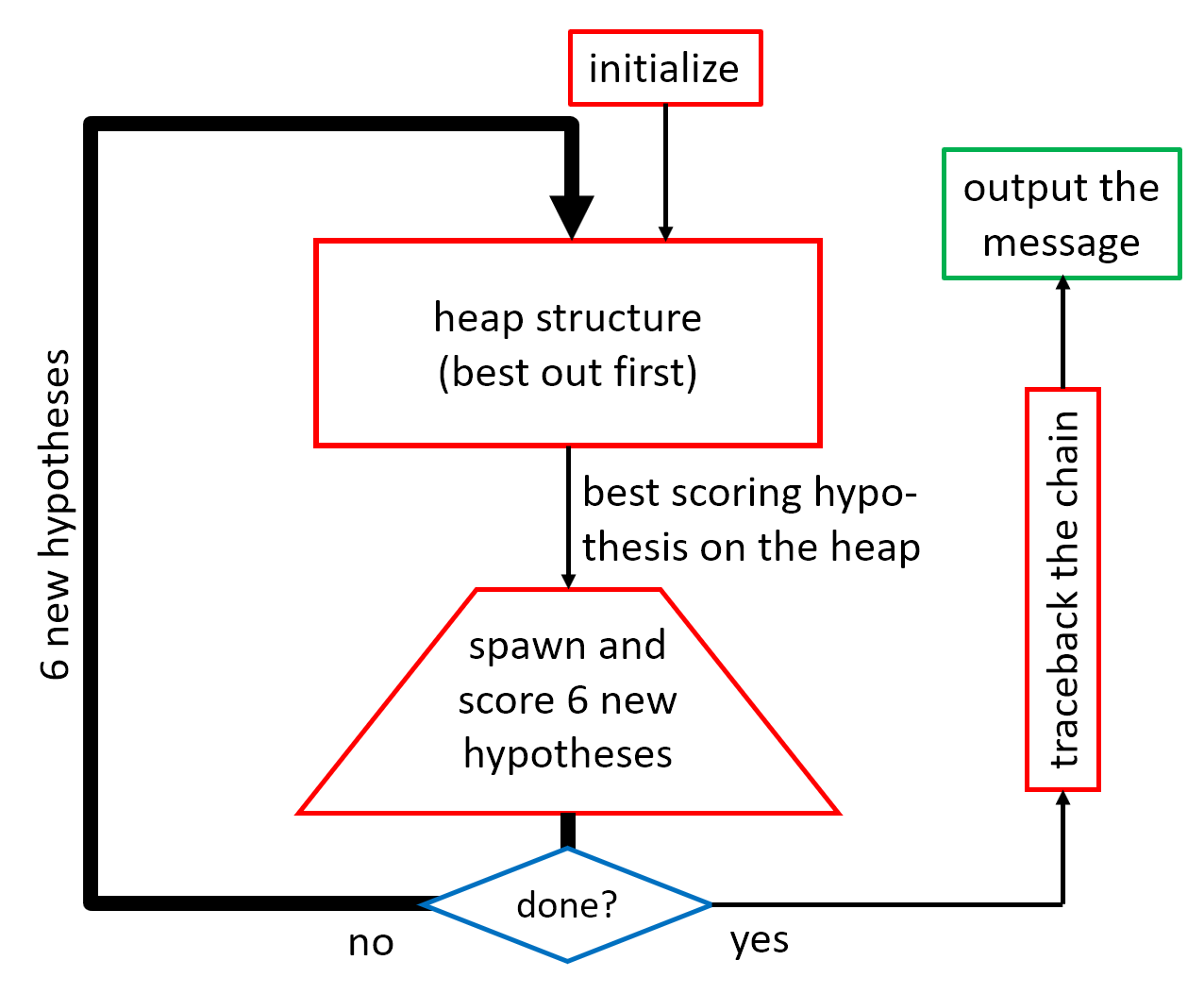}
\caption{The HEDGES decoding algorithm is a greedy search on an
  expanding tree of hypotheses.  Each hypothesis simultaneously
  guesses a message bit $b_i$, its bit position index $i$, and
  its corresponding character position index $k$.  While the tree can in
  principle grow exponentially, a ``greediness parameter'' $\Pok$ (see
  text) limits its growth: Most spawned nodes are never revisited.}
\label{fig3}
\end{figure}

We can formalize the above discussion as follows.  Let
$\HH[i,b_i,B_i,k]$ denote the joint hypothesis that the values
$i,b_i,B_i$ are all correct and synchronize to the observed codestream
character $C_k^\prime$ through equation \eqref{encodeeq}.  As a node
in the search tree, the hypothesis $\HH[i,b_i,B_i,k]$ spawns six child
hypotheses, each of which can be scored with additional penalty
$\Delta P$ (to be added to their common parent's accumulated penalty)
as follows:
\begin{equation}
\begin{array}{ll}
\HH[i+1,\{0,1\},B_{i+1},k]: &\Delta P = \Pdel \\
\HH[i+1,\{0,1\},B_{i+1},k+1]: &\Delta P= (\Pok \text{ if $C=C^\prime$ else } \Psub) \\
\HH[i+1,\{0,1\},B_{i+1},k+2]: &\Delta P= (\Pins+\Pok \text{ if $C=C^\prime$ else } \Pins+\Psub)\\
\end{array}
\end{equation}
Here $\Psub,\Pins,\Pdel$ can be thought of as respectively the log
probability penalties for substitution, insertion, or deletion errors (but
see Section \ref{tradeoffs}).  $\Pok$ is the penalty or, if negative, reward, for an
agreement between the computed and received codestream characters $C$ and $C^\prime$.
In the comparison notated above as $C=C^\prime$, the index of $C$ is the first
parameter in the hypothesis $H$, while the index of $C^\prime$ is the last
parameter in $H$. Note that a child node's $B_{i+1}$ is always computable from its
parent's $B_i$ and $b_i$.

How can we practically search this huge tree?  A conceptual starting
point is the famous A* search algorithm \cite{astar}, a best-first
(that is, ``greedy'') search utilizing a heap data structure. A*
assigns a heuristic cost to every node that is the sum of its actual
cost plus a quantity less than or equal to the smallest possible
additional cost that it can incur in reaching the goal.  (For a tree
of constant depth, this is equivalent to adding a reward
for every step taken closer to the leaf nodes, i.e., a negative
constant $\Pok$ above.)  Figure \ref{fig3} shows the logical flow of
an A* search, and also the HEDGES decode algorithm.  As already
remarked, in coding theory, this kind of decoding strategy is called
``sequential decoding''.

Provably, A* always finds the best path.  For our application,
unfortunately, it is exponentially slow, because actual errors along
the true path cause too many spawned hypotheses to be revisited; and
because its termination criterion is too restrictive, again leading to
too many spawned hypotheses.

To ameliorate these problems we make two heuristic modifications of
A*: First, we allow $\Pok$ to be more negative than that sanctioned by
A* and tune its value heuristically.  While we thus lose the
guarantee of finding exactly the shortest path, we heuristically
encourage the search not to revisit earlier hypotheses after a
sufficiently lengthy run of successes along one particular chain.
Second, we adopt a ``first past the post'' termination criterion.
That is, the first chain of hypotheses to decode the required $L$
bytes of message wins.  It is not obvious (or, by us, provable) that
these heuristics should result in a workable or efficient algorithm,
but we will demonstrate by numerical experiment that it does.

\subsection{Use of Salt to Protect Critical Message}\label{saltsec}
Above, we noted the importance of protecting message bits that
determine the ordering or ``serial number'' of strands for the
outer, concatenated Reed-Solomon code.  In equation \eqref{encodeeq}
(and Figure \ref{fig2}) we allowed for some number of bits of known
salt $S_i$ when message bit $b_i$ is encoded.  Here is how this salt
is enabling of extra protection: Suppose we want to protect an initial $n$ message bits.  Then define recursively
the salt by
\begin{equation}
  \begin{split}
    S_0 &= 0\\
    S_i &= S_{i-1}b_i,\quad i=1,\ldots,n-1\quad\text{(denoting concatenation)}\\
    S_i &= S_{i-1},\quad i\ge n\\
  \end{split}
\end{equation}
Most errors in the first $n$ bits will be corrected as usual by the
shortest-path heap search.  But any residual error that gets through
will ``poison'' the salt for the entire rest of the strand, rendering
it undecodable.  In effect we convert an error in the protected bits
into an erasure of the whole strand.  This may seem drastic, but it is
just what we want: An strand with incorrect serial number (and hence
incorrect ordering among other strands) would look like a
strand of errors (with probablility 255/256 per byte) to the outer
R-S; an erased strand is equivalent to only half as many errors.

\subsection{Code Rates Other than One-Half}\label{othercoderates}
A simple modification of the encode and decode algorithms described in
Sections \ref{encoding} and \ref{decoding} allows for code rates other
than one-half.  Take the input bitstream of expression
\eqref{bitstream} and partition it into a stream of values $v_k$ with
variable numbers of bits in the range $0$ to $2$, according to a
repetitive pattern like the ones shown in Table \ref{tab1}.

\begin{table}[h!]
  \begin{center}
    \begin{tabular}{c|l|c}
      \text{Code Rate} & \text{Pattern} & \text{$\Pok$ (see text)}\\
      \hline
      0.750 & $2,1,2,1,\ldots$ & $-0.035$\\
      0.600 & $2,1,1,1,1,2,1,1,1,1,\ldots$ & $-0.082$\\
      0.500 & $1,1,\ldots$ & $-0.127$\\
      0.333 & $1,1,0,1,1,0,\ldots$ & $-0.229$\\
      0.250 & $1,0,1,0,\ldots$ & $-0.265$\\
      0.166 & $1,0,0,1,0,0,\ldots$ & $-0.324$\\
    \end{tabular}
    \caption{Mapping of bits $b_i$ to variable-bits $v_i$ for various code rates}
    \label{tab1}
  \end{center}
\end{table}

Here are two examples showing how to interpret the entries in Table \ref{tab1}
(with adjacency denoting two-bit values in $\z_4$):
\begin{equation}
  \begin{split}
    \text{Rate 0.750: } v_0 &= b_0b_1, v_1 = b_2, v_2 = b_3b_4, v_3 = b_5,\ldots\\
    \text{Rate 0.250: } v_0 &= b_0, v_1 = 0, v_2 = b_1, v_3 = 0,\ldots\\
  \end{split}
\notag\end{equation}

Equation \eqref{encodeeq} for encoding now becomes
\begin{equation}
C_i = K_i + v_i = F(S_i,I_i,V_i) + v_i \;(\text{mod } 4)
\label{varencodeeq}
\end{equation}
where $V_i$ is composed of concatenated previous variable bits.
Pattern values of 0 provide one bit of additional redundancy check
relative to the base case of code rate one-half, while pattern values
of 2, encoding 2 bits per DNA character, provide one less bit.  By
construction the code rate is one-half the average of the integers in
the pattern.  The column in the table labeled $\Pok$ will be explained
in Section \ref{tradeoffs}.

Decoding follows exactly the same pattern.  Guessing a two-bit $v_i$ spawns
12 child hypotheses, while guessing a zero-bit $v_i$ spawns only 3.

\subsection{Choice of, And Trade-Offs Among, Parameters}\label{tradeoffs}
For encoding, the parameter choices are (i) the choice of code rate and
variable bit pattern (as in Table \ref{tab1}), the default
case being code rate $0.5$; (ii) the number $q>0$ of
low-order bits of position index in the hash; (iii) the number $r>0$ of
previous message bits in the hash; (iv) the number $s\ge 0$ of salt bits;
and (v) the number $n\ge 0$ of initial message
bits to be protected by salt.

It might at first seem that bigger is better for both $q$ and $r$, but
this is not the case.  Restricting $r$ to a smaller value better
allows the heap search to recover from previous errors, basically by
finding an acasual (i.e., ``wrong'') path that coincidentally puts it
back on track.  As for $q$, restricting it to a smaller value could be
useful in case one desires the capability of jumping into the middle
of an undecoded message: The heap can then be initialized with all
possible values of $I$ and $B$ (cf.~Figure \ref{fig2}).  For our
system design, Section \ref{systemdesign}, this is not a necessary, or
useful, capability, however. For the baseline validation experiments
in Section \ref{validation}, we take $q=10$, $r=8$, $n=16$ or $24$.

For decoding, we need to know the encoding parameters, and must now
also choose values for $\Psub,\Pdel,\Pins$, and $\Pok$.  While,
conceptually, these are negative log probabilities of the occurrence
of the different kinds of errors (which can be known only after the
fact), we adopt a more empirical approach.  First, we take
$\Psub=\Pdel=\Pins$ to give the HEDGES decoding algorithm equal
robustness against all three kinds of errors.  Second, we note that
the search for shortest path is invariant under applying the same
linear (or affine) transformation to all four $P$'s.  So, without loss
of generality, we may take $\Psub=\Pdel=\Pins=1$, leaving $\Pok$ as
the only free parameter.  We determine optimal (or at least good)
values for $\Pok$ by numerical experiment.  We find that the optimal
$\Pok$ depends only negligibly on the encoding parameters $q$ and $r$,
and only slightly on the length $L$ of the strand, but it
does depend on the code rate.  Good values for various code rates are
given in the third column of Table \ref{tab1}.

Implicitly, the choice of $\Pok$ reflects a tradeoff between
computational workload and decode failure probability.  $\Pok$ that is
too negative results in too greedy a search, which is fast but can get
stuck in a blind alley that requires us to declare the rest of the
strand as an erasure (hence its dependence on strand length).  On the
other hand, $\Pok$ that is insufficiently negative results in a too
large, potentially exponential, expansion of the size of the heap.
Happily, there is an accessible range of workable values.  Changes of
$\sim 10$\% in $\Pok$ matter little, and our values are implicitly
tuned for best performance on strand lengths in the range $\sim 100$
to $\sim 1000$.

In Section \ref{decoding} above we limited the guesses for $\Delta$ to
only $\{-1,0,1\}$ so as to limit the expansion of the heap.  This
results in more than one consecutive insertion or deletion being
improperly scored.  For example, without the possibility of skew
$\Delta=-2$, the shortest available path through two deletions $\ldots
DD \ldots$ declares a spurious substitution $\ldots DSD \ldots$.  In
practice, this makes little difference, because double deletions are
significantly less common than single deletions, and because other,
completely incorrect, paths score much worse.

It is an important point that choosing any set of decode parameters is
not an irrevocable choice.  Given a DNA message, one can make multiple
tries, varying the decode parameters adaptively until acceptable
performance is achieved. One can evaluate success by running time and
by the count of errors needing correction by the outer R-S code. The
parameter values that we suggest may be viewed as starting points.

\section{Computer Validation Experiments}\label{validation}
We have implemented HEDGES in C++ code, with also a Python interface
for convenience.  (We similarly implemented a compatible Python
interface to the published ``Schifra'' implementation of
Reed-Solomon.\cite{schifra}) For tests on individual strands of length
$L$, we encode a random stream of message bits and degrade
the resulting codestream by errors with a specified Poisson-random
total rate, divided equally among the three error types, substitution,
insertion, and deletion.  Unless otherwise stated the HEDGES code
rate is one-half.

\begin{figure}[thb]
\centering
\includegraphics[width=4in]{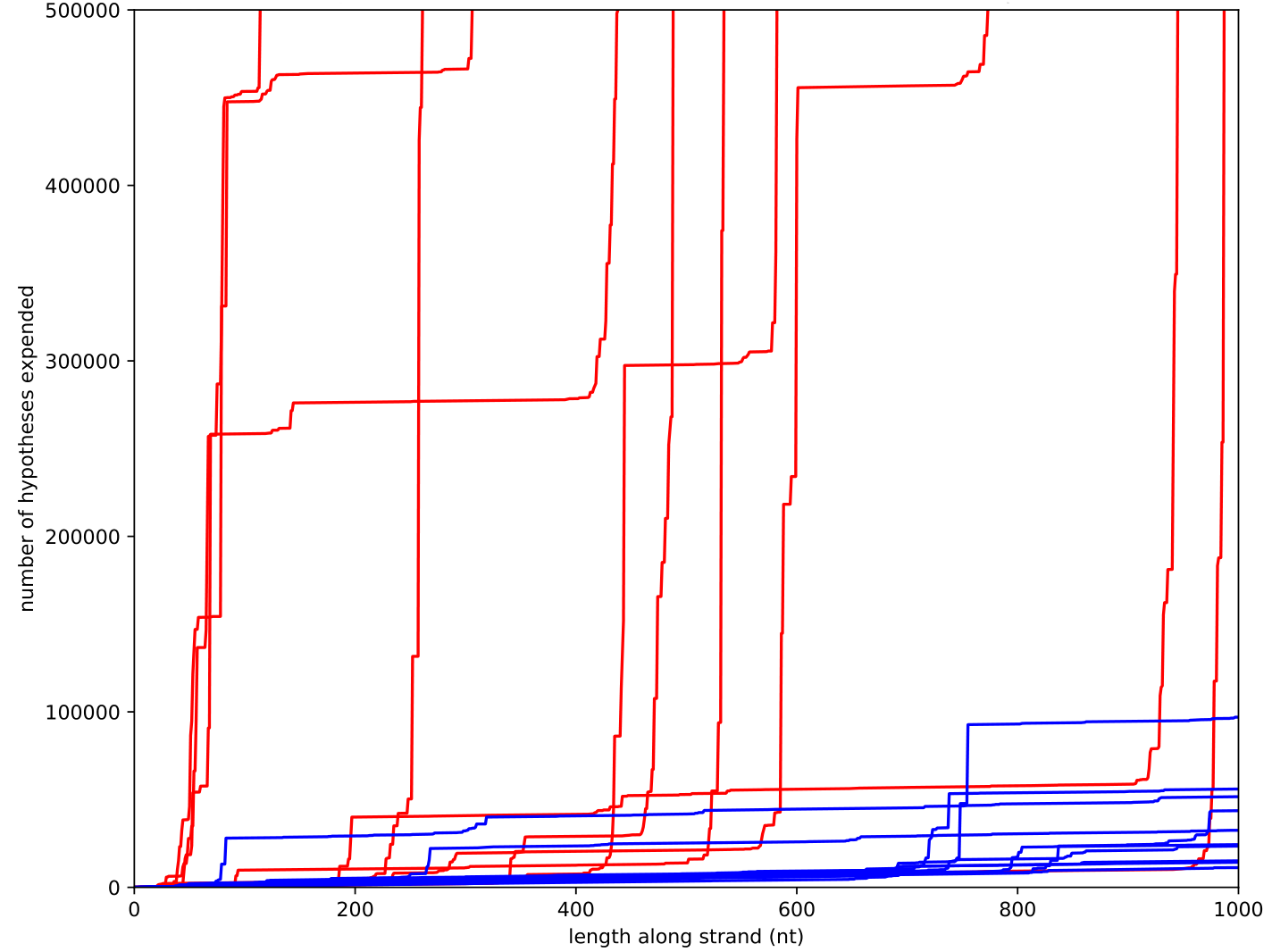}
\caption{Hypotheses expended (heap size) in decoding a long strand
  with 5\% DNA character errors.  Ten examples are shown each of
  successful decodes (blue) and unsuccessful decodes
  (red). Unsuccessful decodes are declared as erasures.  Although here
  shown equally, red cases are actually much rarer than the blue (see also
  Figure \ref{fig5}).}
\label{fig4}
\end{figure}

We allow each decode a ``hypothesis budget'', that is, a maximum size
to which the heap is allowed to expand.  If, along a strand, a
decode exceeds its budget, we declare subsequent message bits in that
strand to be erasures.  Figure \ref{fig4} shows examples of how decodes
expend their budget along a long strand.  There is a sharp bifurcation
between decodable strands, which typically expend $\lesssim 100$
hypotheses per decoded bit, and undecodable strands, which go into
blind alleys and readily expend $\gtrsim 1000$ hypotheses per decoded
bit.  The figure shows 10 selected examples of each type.  In practice
undecodable strands are much rarer than decodable ones.

\begin{figure}[thb]
\centering
\includegraphics[width=5in]{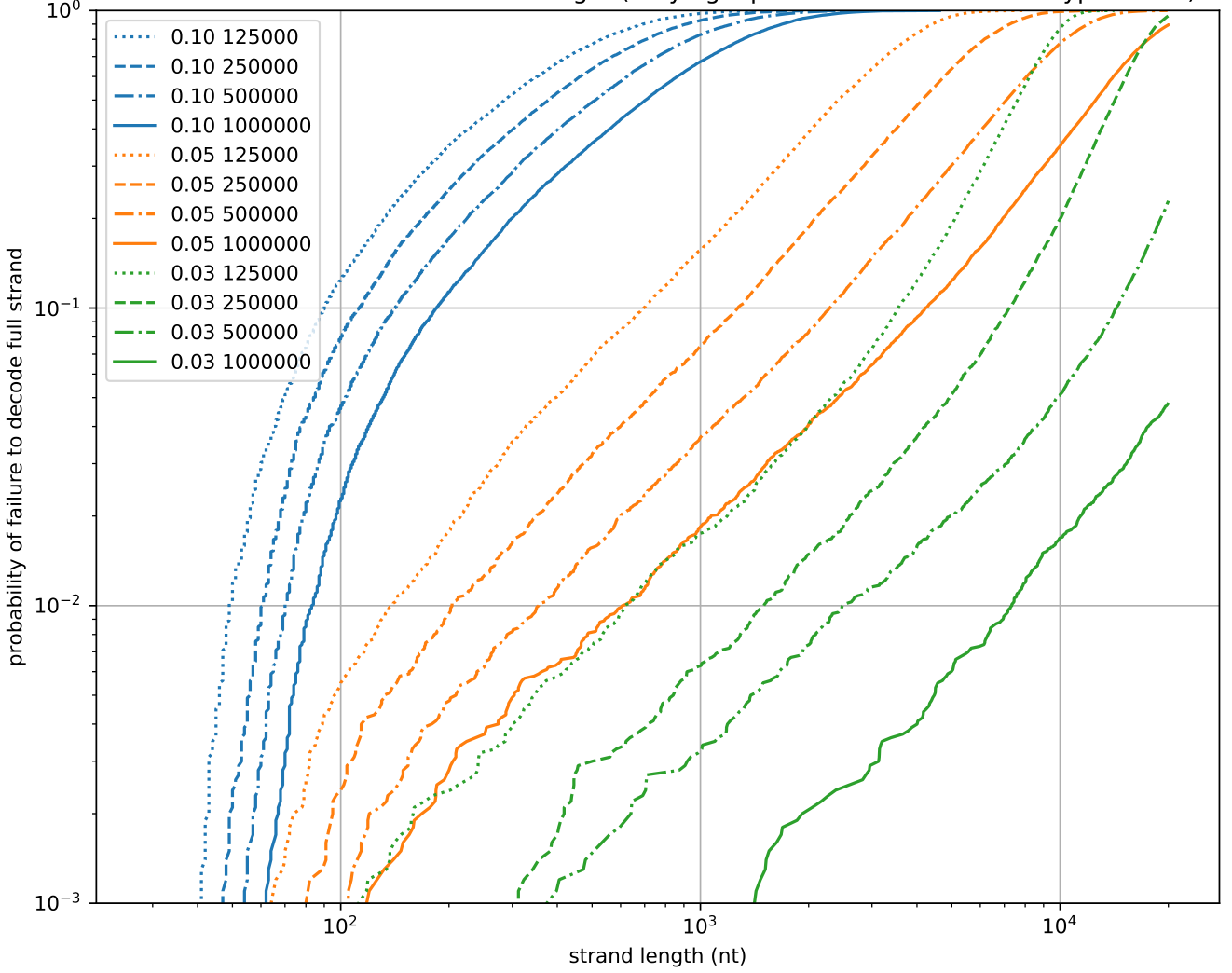}
\caption{HEDGES decode failure rates vs. strand length of a half-rate
  code as a function of total DNA error rate (3\%, 5\%, 10\%) and
  hypothesis budget ($1.25\times 10^5$, $2.5\times 10^5$, $5\times
  10^5$, $1\times 10^6$).  Failure rates $\lesssim 10^{-2}$ are fully
  correctable by the concatenated Reed-Solomon code.}
\label{fig5}
\end{figure}

\begin{figure}[thb]
\centering
\includegraphics[width=6in]{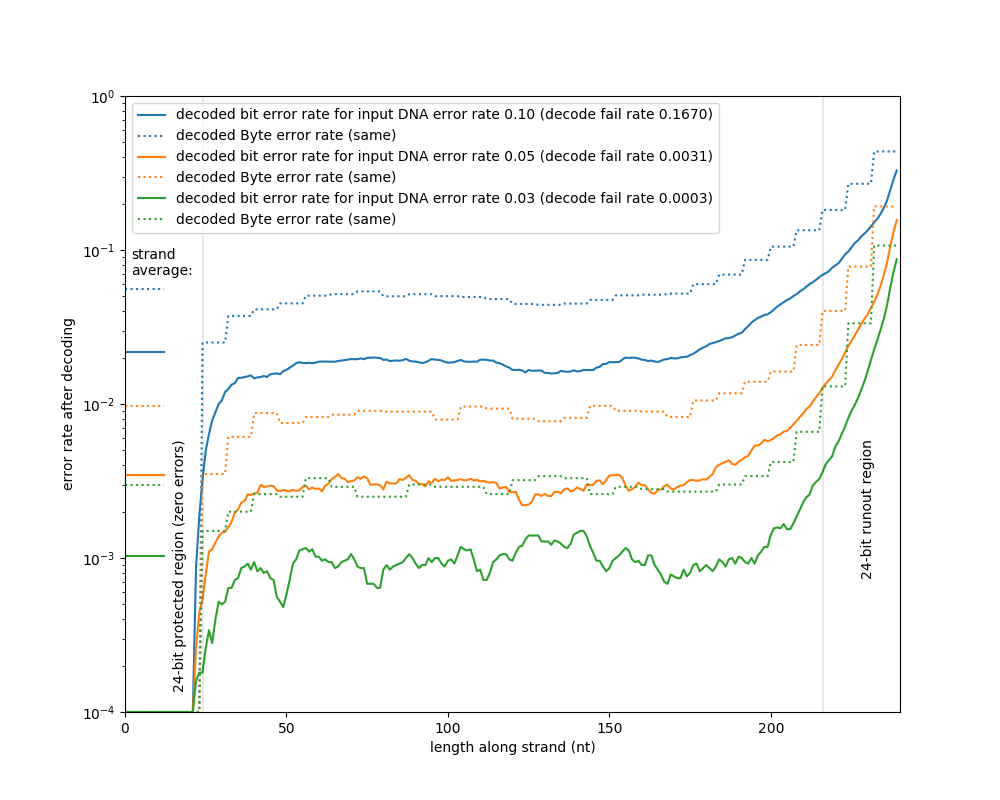}
\caption{Remaining uncorrected substitution error rates after HEDGES
  decoding, excluding the erasures implied by Figure \ref{fig5}.  For
  this numerical experiment, strands of length 240 bases were encoded
  with a half-rate code, with 24 initial bits protected by salt (see
  Section \ref{saltsec}).  Strand averages (exclusive of protected and
  runout regions) are shown at the left edge of the figure. Error
  rates rise in the runout region, because fewer downstream bits of
  redundancy are available.  Strand average Byte error rates $\lesssim
  10^{-2}$ are readily corrected by the outer Reed-Solomon code.}
\label{fig6}
\end{figure}

Figure \ref{fig5} shows decode failure rates actually achieved by a
half-rate code, as a function of length of strand, input total error
rate, and hypothesis budget.  One sees that, for strand lengths in the
useful range 100--1000, failure rates $\lesssim 10^{-2}$ are readily
achievable for total input error rates up to $\sim 5$\%.  For an input
error rate of 3\%, strand lengths up to $10^4$ are feasible.  Failure
rates $\lesssim 10^{-2}$ are easily absorbed as erasures in the error
budget of an outer, interleaved Reed-Solomon code (see below, this
section).

In the case of a successful decode, there may remain uncorrected
substitution errors.  Figure \ref{fig6} shows the uncorrected (output)
bit, and byte, error rates along strands of length 240 for the three
input codestream character error rates 3\%, 5\%, and 10\%.  The
uncorrected error rates vary along the strand for two reasons: First,
for this experiment, we applied salt protection to the first 24
message bits (that is, 24 characters for the half-rate code).  One
sees that this worked as advertised: There were no uncorrected errors
in the first 24 bits.  Second, uncorrected error rates are seen to
rise as the length of the strand is approached.  Although undesirable,
this is an inevitable feature of HEDGES.  As the strand end is
approached, there are fewer redundancy checks available downstream,
making the greedy search algorithm less selective.  Our system design
(Section \ref{systemdesign}) allows for specifying some number of
``runout'' bits at the ends of the strands, encoding zeros and not
part of the message packet.  How much runout to allow depends on how
much one wants to burden the R-S error budget.  For this numerical
experiment, we assumed 24 runout bits.

It is notable that the byte error rates in Figure \ref{fig6} are only
$\approx 3$ times the bit error rates, rather than $\lesssim 8$ times
(depending on the bit error rate) if the errors were randomly
distributed.  This shows that HEDGES's uncorrected errors are bursty,
which is good for input to R-S and gains some overall efficiency.

Exclusive of the salt-protected and runout regions, the uncorrected
bit error rates for this experiment are about $1\times10^{-3}$,
$3.5\times10^{-3}$, $2\times10^{-2}$ for input error rates 3\%, 5\%, and
10\%, respectively.  The corresponding byte error rates are
$3\times10^{-3}$, $1\times10^{-2}$, and $6\times10^{-2}$.

What level of uncorrected errors may we allow to get
through to the R-S outer code, with a high probability that they
will there be corrected?  $RS(255,223)$ is able to correct 16
byte-errors, or any combination of byte errors and erasures
whose equivalent number is
\begin{equation}
N_{\text{equiv}}=(\text{byte errors}) + 0.5\times(\text{byte erasures}) \le 16
\notag\end{equation}
The probability of failure to completely correct 255 bytes, for Poisson
random byte errors/erasures is thus the cumulative probability
\begin{equation}
P_{\text{failure}} = \text{PoissonCDF}(k>16\,|\,\lambda=255 P_{\text{equiv}})
\label{Poisson}\end{equation}
where $P_{\text{equiv}}$ is the byte error rate (B.e.r.) plus half the
erasure rate.  As mentioned above, we interleave and apply the R-S
code diagonally (see Figure \ref{fig1}) because (i) strands may be
missing, (ii) byte errors along a single strand may be bursty, and
(iii) it is found experimentally that sequencing and synthesis error
rates can be different (often larger) near to the end of strands.  The
interleaved diagonal pattern ensures that no single $255$ length R-S
packet gets handicapped by an error rate much different than the
average across the whole strand, and that its number of errors will be
distributed with (close to) Poisson statistics.  Table \ref{tab2}
evaluates equation \eqref{Poisson} for relevant values of
$P_{\text{equiv}}$.  One sees that a value $P_{\text{equiv}} \lesssim
1\%$ are adequate to guarantee error-free decoding of messages of
gigabyte length or longer.

\begin{figure}[thb]
\centering
\includegraphics[width=5in]{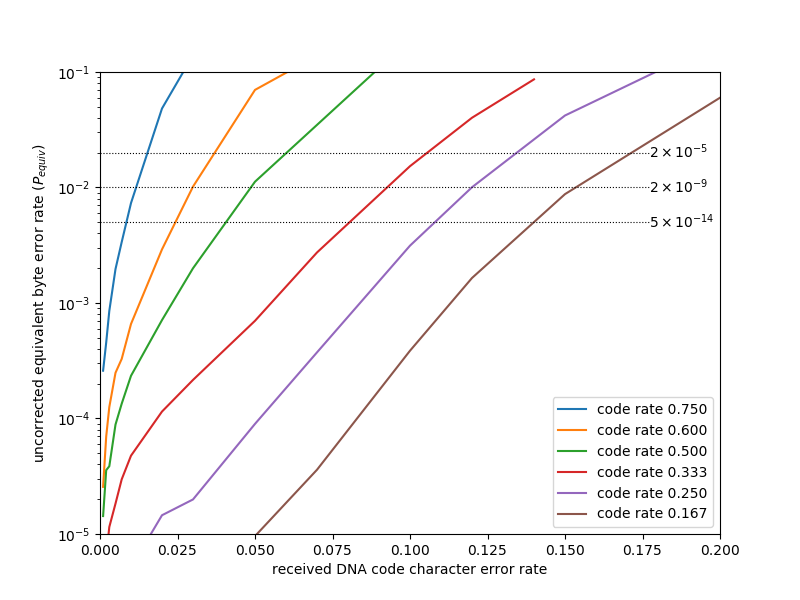}
\caption{Effective rate of uncorrected byte errors after HEDGES
  decoding for 6 code rates.  The curves give values
  $P_{\text{equiv}}$ that, after correction by the outer Reed-Solomon
  code, give the error rates labeled on the horizontal dotted lines
  (values from Table \ref{tab2}).  Intersections with the dotted lines
  correspond to final error rates (post Reed-Solomon) shown on the
  right.  This numerical experiment is done with strand length
  $L=300$, no salt protection, and 2 bytes of runout (see text).}
\label{fig7}
\end{figure}

\begin{table}[h!]
  \begin{center}
    \begin{tabular}{l|l}
      $P_{\text{equiv}}$ & $P_{\text{failure}}$\\
      \hline
      0.005 & $5.25\times 10^{-14}$\\
      0.010 & $2.08\times 10^{-9}$\\
      0.020 & $2.53\times 10^{-5}$\\
      0.030 & $2.39\times 10^{-3}$\\
      0.040 & $3.16\times 10^{-2}$\\
    \end{tabular}
    \caption{Probability that the outer code $RS(255,223)$ fails to
      correct 100\% of message errors as a function of $\Peq$, the
      inner-code HEDGES output byte error rate (plus half the erasure
      rate). One sees that $\Peq < 0.01$ is sufficient for error-free
      decoding of gigabyte-length messages.}
    \label{tab2}
  \end{center}
\end{table}

Figure \ref{fig7} now shows the results of a numerical experiment
evaluating $P_{\text{equiv}}$ as a function of input DNA error rates
for six different code rates.  The evaluation was done with strand
length $L=300$ (as a variant of the value $L=240$ in Figure
\ref{fig6}), no salt protection, and 2 bytes of runout on each strand
(errors in which are not counted).  Using Table \ref{tab2}, one sees
that a DNA error rate of about 1\% is correctable at code rate 3/4
with probability effectively 1, increasing to a correctable error rate
of about 15\% at code rate 1/6.

\FloatBarrier

\section{Channel Capacity re Shannon Limit}

We might wonder how close the results of Figure \ref{fig7} come to the
absolute bound of the Shannon limiting channel capacity \cite{sh}.
Unfortunately, computing the Shannon limit for even the simplest case
of a binary deletion channel, let alone channels with also
insertions and substitutions, remains a difficult unsolved
problem \cite{mitz, radu}. Still, it is possible to get some idea by
making an informed estimate as follows.

\begin{figure}[thb]
\centering
\includegraphics[width=5in]{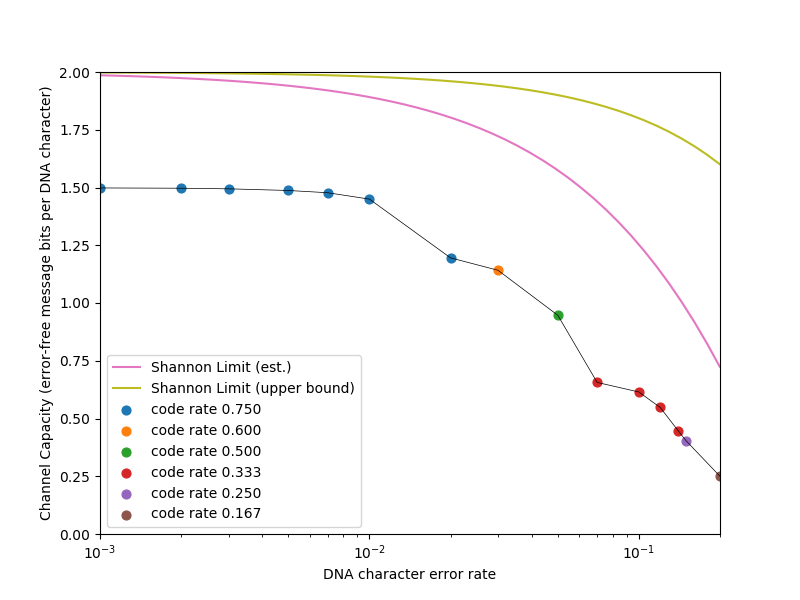}
\caption{Comparison of the channel capacity achieved with HEDGES to an
  estimate of the Shannon limiting capacity.  The dots show the
  greatest of the capacities achieved by the various code rates in
  Table \ref{tab1}.  As expected, the smaller code rates become
  optimal for the larger input error rates. Codes with $r > 0.75$ (not
  tried) would give even better performance at the smallest
  error rates (left side of of the graph).}
\label{fig8}
\end{figure}

A remarkable theorem of Shannon
\cite{fbchan} proves that the channel capacity of a forward error-corrected
channel is identical to that of a ``feedback channel,'' where the sender
gets to see (error-free) what was actually received, and then send
correcting information {\em post hoc}.  We can thus estimate channel
capacity by reducing the maximum capacity (for DNA, 2 bits per character)
by the entropy of the necessary correction messages.  The reason that
this is an estimate only (strictly, a lower bound), is that we may not
be sending the optimally short correction messages, especially as the
error rate becomes large.

In our case, suppose $p$ is the character error rate.  As before,
assume equal probabilities $p/3$ for the three kinds of errors.  Then
we can estimate the Shannon limit of the channel as
\begin{equation}
  C_{\text{Shannon}} = 2 - [H_2(p) + p\log_23 + \tfrac{1}{3}p\log_24 +
    \tfrac{1}{3}p\log_23]
\label{cest}\end{equation}
where $H_2$ is the entropy of a binary choice,
\begin{equation}
H_2(p) \equiv -p\log_2 p - (1-p)\log_2(1-p)
\end{equation}
In equation \eqref{cest}, the first term in brackets is the cost of
communicating whether a particular code character marks the position
of an error. The second term is the cost of telling which kind of
error (substitution, deletion, insertion).  The third term is the cost
of communicating the missing character in a deletion.  The fourth is
the similar cost for a substitution.  While strictly only a lower
bound, analogous results for binary deletion channels \cite{radu}
suggest that equation \eqref{cest} is actually a good
approximation for small values of $p$.

We now calculate the channel capacity actually achievable with HEDGES
by the relation
\begin{equation}
C_{\text{HEDGES}} = 2 \times (\text{code rate}) \times [1-H_2(\peq)]
\end{equation}
Here the factor 2 is the number of bits per DNA character, while the factor in
square brackets reflects the loss of channel capacity to an (assumed perfect)
concatenated outer code that corrects all of HEDGES' uncorrected bit errors.

Figure \ref{fig8} shows the results of the comparison.  One sees that HEDGES
achieves a respectable fraction, $\gtrsim 0.5$, of the estimated Shannon limit
for DNA character error rates up to 20\%.

\section{Discussion}
Previous work on DNA information storage, despite the increasing
sophistication of methods, has largely ignored the possibility of
directly correcting insertion and deletion errors by an appropriate
error-correcting code.  Instead, most previous work has relied on
multiple sequence alignment after sequencing DNA messages to
significant depths.  In effect, though not always acknowledged, this
method is an inefficient multiple-repetition code.

This paper developed a coding technique, termed HEDGES, for the direct
correction of insertions and deletions, along with substitutions,
workable with (combined) DNA character error rates up to 20\%, and at
a respectable fraction of the Shannon information limit.  The code,
HEDGES, was optimized for use as the inner code in an overall design
with an outer concatenated code that will generally be interleaved
across DNA strands.  Used with HEDGES, the outer code need not be
indel-aware and can be a conventional ECC like Reed-Solomon.

\subsection*{Acknowledgments}
We have benefitted from communication with Dave Forney, Dan Costello,
and Vince Poor, and from our continuing collaboration in related
matters with Ilya Finkelstein and Stephen Jones.

\FloatBarrier

\end{document}